**Design of 2-m to 6-m Liquid Mirror Containers.**

Robert Content

University of Durham, Astronomical Instrumentation Group, Department of Physics, South Road, Durham, DH1 3LE, UK

Subject headings: Telescopes, Instrumentation, Liquid Mirrors

**Abstract**

A new design is proposed for large (up to 6-m) liquid mirror containers. The design uses Kevlar, foam and aluminum, as in previous designs, but with a different configuration that makes the container lighter, stronger and more rigid. The results of finite element analysis are presented, consisting in the deformations due to temperature changes and to weight, and in the security factor for each material when maximum constraints are applied. Tilt rigidity is also analyzed. They show that the composite material construction technique gives a good performance up to 6 meter diameters. The figures and tables contained in this paper can be used as recipes to build containers having diameters between 2 and 6 meters.

**1. Introduction**

The technology of liquid mirrors has now been fully demonstrated in the laboratory (Borra et al. 1992; Borra, Content & Girard 1993; Girard & Borra 1997; Tremblay & Borra 2000) and in the field (Borra et al. 1988; Hickson et al. 1994; Sica et al. 1995; Wuerker 1997; Hickson & Mulrooney 1998) for 1-m to 4-m class systems. By making available for the first time large optical quality optics at low cost, it increases the performances in many domains of sciences. The most complete facility to date is the NASA Orbital Debris Observatory (NODO) which uses a 3-m liquid mirror (Potter & Mulrooney 1997; Mulrooney 1998). This telescope has also been used to make astronomical research (Content et al. 1989; Cabanac, Borra & Beauchemin 1998) that proves the large scientific return for a given cost and the feasability of long astronomical programs using liquid mirror observatories.

Liquid mirror is the technology that gives the largest étendu (product of primary surface by field solid angle: the $A\Omega$ product) per dollar. Because of its low cost, the next logic step is the building of observatories containing a large number of 4-m to 6-m independant liquid mirror telescopes in the same building, all with their own detectors and observing the same field. For the price of one stearable 4-m to 6-m telescope, the large number of telescopes will give the light gathering power of a much larger mirror and with a much larger field of view than such a large mirror can achieve at a reasonable cost. The present article describes a rigid and low cost design for the containers of these observatories.

The main early work of finite element analysis of containers is the unpublished M.Sc. thesis of Arrien (1992). The main conclusion of this work was that composite containers can be as large as 6-m in diameter but that larger containers would be too heavy and expensive. This conclusion may not be valid for the new design proposed in this paper since the container weight is much smaller than what was studied by Arrien (1992). Still, the present study is limited to containers up to 6-m in diameter. This is mostly because the cost of the field corrector rapidly increases with the mirror diameter, so it is probably less expensive to build a large number of 4-m to 6-m mirrors with large fields of view than a large liquid mirror with the same total collecting surface and field of view.

The materials in the container design of the present article are the same than in the container design of Hickson, Gibson & Hogg (1993). All present liquid mirrors in use in the world use the basic design developped at Laval University with a container using, in most case, the kevlar and foam and aluminium design described by Hickson, Gibson & Hogg (1993).

Composite containers are not necessarily the best design to adopt. The space frame design used by Hickson et al. (1998) probably is a better one for mirrors having diameters larger than 6 meters. However, composite containers have the important advantage that materials are inexpensive and that they are easier to build since they can be constructed with limited technical resources and unskilled inexpensive labor (e.g. graduate students). This is a major consideration, for liquid mirrors only make sense if they are considerably less expensive than conventional glass mirrors so that, for the same price, the huge light collecting power compensates the inhability to point.

In the present article, building on the previous studies, a detailed engineering study of the containers is made. The previous designs had some limitations that make them difficult to scale to larger sizes. The new design is lighter, more resistant and more rigid. Finite element computations of containers under load based on this design are presented. The important temperature effects on the containers are also described. The goal is to produce practical recipes that will allow one to build a working LMT container without resorting to his own finite element computations. Measurements on actual containers are presented in a companion paper (Borra et al. 2003). They confirm the validity of the calculations in the present paper.

A complementary introduction is given in the companion paper.

## 2. Generic description of a liquid mirror system

A complete description of a liquid mirror system can be found in the companion paper. In the present article, only the basic components are reminded to the reader. The liquid mirror assembly is made of 3 parts: the rotation system, the container and the mirror itself.

The rotation system rotates the container and precisely maintain the rotationnal speed. It is composed of a high precision turntable (usually an air bearing), a motor and an electronic system to precisely control the motor speed. Also part of the system is the drive, which can be direct or through a system of 2 pulleys and a belt.

The mirror is obviously made of mercury but should also include what is directly linked to the liquid mirror technology, so it includes the components that give the high surface precision and the small thickness of mercury. The high surface precision is obtained by spincasting resin on the top of the container at the speed of use (Fig. 1, top). The bottom of the mercury layer has then the same parabolic shape than the mercury surface giving a uniform thickness of liquid. The small liquid thickness is obtained by making a groove in the resin at the edge of the mirror and pumping the mercury in the center while the mirror is rotating. The groove controls the effect of the mercury surface tensions that would otherwise brake the mercury surface at a thickness of less than 4 mm. Mercury, resin, groove and the assembly in the center to pump the mercury therefore make the mirror.

The link between the turntable and the mirror is the container, which is the subject of this paper. It must be rigid but also light to reduce the cost of the bearing. This is achieved by using composite materials attached in the center to an aluminium structure. The first containers we made had a flat top but, to minimize the weight of the spincasted resin, the top of the container must have the same parabolic shape than the mercury.

## 3. Description of the design

In the simplest design of a liquid mirror container used with the early mirrors (Borra et al. 1992), a disk of foam would be covered on each side by a layer of Kevlar with an aluminum plate the size of the bearing attached to the bottom center (Fig. 1, middle top). While slightly different, the design for the existing 3-m class mirrors in use is not far from that design (Hickson et al. 1994) and is subject to the same limitations (Fig. 1, middle bottom). In the new design (Fig. 1, bottom), an aluminum cylinder and a second aluminum plate have been added in the centre, the bottom of the container has a conical shape with the top and bottom layers of Kevlar joining at the edge, and the thickness of Kevlar increases in step from the edge to the centre. While conical bottoms (Fig. 1, middle bottom) and variable thicknesses of Kevlar have been used before, it was at a much smaller level than in the new design. Small aluminium cylinders have also been used but they served primarily as references for position measurements, not for strengthening the containers, a task needing much larger cylinders.

### 3.1. The aluminum cylinder

In the simplest design, most of the weight of the liquid mirror is supported by a small surface of foam near the edge of the aluminum plate. The major inconvenient of this design is that the foam endures severe stresses with a risk of large deformations for large (so heavy) mirrors. By placing an aluminum cylinder in the foam at the centre (Fig. 1, bottom) and gluing foam and cylinder together, the foam is sustained by the whole of the surface of the cylinder. With a larger surface to support the weight, the stress in the foam is much smaller. The cylinder is also useful for increasing the rigidity in tilt. This is critical since the container is unstable if the rigidity is too small (Content 1992; this paper appendix 2). The problem comes from the fact that, being liquid, mercury can move. If a tilt is forced upon the container, the mercury flows down the slope. This in turn generates a moment that tilts the container further. For a given tilt, if the moment created by the unbalanced mercury is larger than the counter-moment due to the rigidity in tilt, the container is unstable with catastrophic consequences.

The different parameters of the cylinder are determined differently. The diameter is made similar to the bearing diameter which is the maximum diameter with a direct support of the cylinder by the bearing. The thickness of the cylinder wall is mostly determined through the optimization process by the rigidity to weight ratio of the container. Increasing the thickness increases the rigidity but also the weight. At constant weight, something must be removed when increasing the thickness, which reduces the rigidity somewhere else. There is therefore a thickness value that maximizes the rigidity. The height of the cylinder, while also determined by the optimization process, is mostly influenced by the total weight we decided for the container.

### 3.2. The Kevlar layers

Computer simulations show that, for a constant total weight, the steeper the conical shape of the bottom surface the larger the container rigidity, with a maximum when the edge thickness of foam is zero. Although prima facie it seems trivial, this zero-thickness edge is a very important feature of the new design (Fig. 1, bottom) and is the shape adopted in the computations. The main physical reason for this choice of shape comes from the low rigidity of the foam, which is 1000 times less rigid than Kevlar. When the 2 surfaces of Kevlar are parallels, as in the simplest design, the foam can sustain large shear deformations. This kind of deformation cannot be reduced by adding kevlar on the top and bottom as would with flexure deformations (Fig. 2).

One problem with this conical shape is that the center of the container contracts more than the edge when the temperature decreases if the top and bottom kevlar surfaces are similar. This is because foam and aluminium have much larger coefficients of thermal expansion than the kevlar. This deformation of the container would break the mercury surface. To avoid this, the top surface is made with more kevlar layers than the bottom. When the temperature decreases, the bottom will then contract more since it is thinner. With the correct kevlar thicknesses, the edge and centre of the container will contract at the same rate with decreasing temperatures. A minimum of 4 layers with different azimutal directions is however necessary to makes the rigidity sufficiently uniforms azimutally.

The variation of rigidity and thermal expansion coefficient with the azimutal angle is due to the direction of the filament in the fabric. We use a fabric with interleaved filaments at 90°, so there is a symetry when rotated by that angle. In the direction of the kevlar filaments, the rigidity is large, but not at 45° where it is about 3 times smaller. This value was calculated with a model of the fabric that used the measured elasticity coefficients of the manufacturer. Even with 2 layers at 45° there is still a ratio of 1.6 between the rigidity in the filament direction and at 22.5°. We therefore use a minimum of 4 layers with azimutal directions in step of 22.5°.

The constraints sustained by the Kevlar increase from the edge to the centre and, therefore, so do the stresses and strains if the layer has a constant thickness. By decreasing the Kevlar thickness near the edge and increasing it near the centre, the deformations and constraints can be reduced while maintaining the same total volume of Kevlar. This was done in the new design by increasing the thickness of kevlar in steps of 4 layers at a series of diameters along the way from the edge to the center of the container (Fig. 3). These diameters were chosen to optimize the mechanical behavior of the container under stress. They were mostly determined through the optimization process by the maximum constraints in the 3 materials, by the deformation of the top surface, and by the rigidity of the container. The result is a series of kevlar steps near the center of the container when optimizing for a minimum temperature of -20°C. A different distribution of kevlar would create larger maximum constraints in the 3 materials. The steps extend further away from the center if we optimize for a minimum of -10°C as shown in Appendix 1 for a 6-m mirror.

At the junction between Kevlar and Aluminium, half the layers of Kevlar are glued on the top of the Aluminium plate, the other half on the bottom. This seems logic to avoid a concentration of temperature induced constraints on the glue at the edge of each plate by doubling the contact surface near that edge. However, this configuration may not be the best. All the layers can also be glued to the top or the bottom of the plate. Most of the containers in use today have the layers glued on the outside of the plate (fig. 1, bottom of centre). No breakage have been recorded so far even after many years of use. More theoretical and test work would be necessary to resolve this question.

## 3.3. Aluminum plates

Two aluminium plates are attached at each end of the cylinder. Their purpose are multiple. They are the link between the kevlar and the cylinder. They also reinforce the center top and bottom surfaces which support the whole weight of the mirror. Their main role however is to reduce the maximum constraint in the foam by avoiding the "squeeze" at the limit between the cylinder and the top and bottom surfaces when the temperature goes down. This happen because the foam contracts more than Aluminium and because of the inhability of kevlar to resist a force in the direction perpendicular to its surface. Without the plates, the foam becomes pinched between the kevlar and the cylinder where they are attached.

## 4. Computer simulations

Simulations of the mechanical properties of the containers discussed in this paper were performed using the ANSYS program version 5.0 of finite element analysis from Swamson Analysis System. The program resolves the equations of static mechanics for a model of an object, here a liquid mirror container, using some approximations and gives the stresses and deformations under the applied loads. These are the weight induced loads and the temperature changes. The main approximation used in the program is to partition the object into smaller objects called elements where the deformations are modeled following a simple equation, for example a second degree polynomial in 3 perpendicular directions. The precision increases with the number of elements but, unfortunately, so does computing time.

## 4.1. Computer model

In order to limit computing time to reasonable values, an axisymetric model was used. It reduces computing time by using the azimuthal symmetry of the liquid mirror to reduce a 3 dimensional problem to a 2 dimensional one (Fig. 4). However, the program can still resolve a fully 3 dimensional problem thanks to its ability to support non-axisymetric loads. In the azimuthal direction, loads can be expanded into Fourier series. The program permits to solve each term of the Fourier series separately because a load that follow a sine form in the azimuthal direction gives deformations and constraints that also follow a sine in the azimuthal direction with the same spatial wavelength. The total deformations and stresses in three dimensions are then obtained by adding together the individual solutions.

Four classes of problems have been solved using the ANSYS program:

1) The stresses and the deformations when a uniform layer of mercury is distributed over the surface. Only gravitational and centrifugal forces are applied to the mirror. This is an axisymetric problem, so only one computation is required.

2) The stresses and the deformations when the temperature is reduced by 40°C. This is also an axisymetric problem.

3) The deformations when a tilt is introduced. The mercury moves down the slope, unbalancing the container and further increasing the slope. Stability requires that this increase be smaller than the original tilt of the container (appendix 2). The load due to the change in position of the mercury follows a sine function in the azimuthal direction with a maximum increase in weight at the bottom of the slope and a maximum decrease (a negative value) at the top. This is a Fourier term solved by the ANSYS program with one computation only.

4) The deformations due to a local force on the non-rotating container, to simulate a bottle of mercury placed there. This is to simulate experiments carried out in the laboratory. The main effect is to tilt the container, giving a measurement of the rigidity in tilt. We also studied the effect of putting two bottles on the container, one at each end of a diameter (Borra et al. 2003). This is a non-axisymetric problem involving the decomposition in a Fourier series in the azimuthal direction of a local load, then finding the deformations due to each term separately as discussed in the preceding paragraph.

## 4.2. Method of optimization

Several different containers were designed using the ANSYS program and their mechanical properties investigated. Most of the work went into designing 2-m to 6-m containers F/1.5 capable of sustaining temperatures as small as -20°C. A 2-m container F/3.5 was also designed for the "Centre Spatial de Liège" in Belgium. Two other containers were designed for the Liquid Mirror Laboratory of Université Laval and built there: A 3.7-m with 3.6-m clear aperture F/1.2, which is now operational and has been toroughtly investigated (Tremblay & Borra 2000), and a 1-m flat container especially built for studying the effects of temperature changes on the containers. These containers give important tests of the reliability of the finite element computations and are described in details in the companion paper.

The design proposed in this paper is for mirrors having a 2-m to 6-m clear aperture. The container is built at room temperature since the polyurethane resins used for this purpose are designed to polymerize at room temperature. However, the container is destined to be used in an astronomical observatory and hence must operate at temperatures down to -20°C without significant deformations and excessive stresses. Since we want to work with a mercury thickness no larger than 1 mm, it is required that the surface of the container deviates from the ideal parabola by less than 1 mm. Figure 4 shows the finite element model used in the calculations. The boundaries of each element are visible. In order to increase the precision of the computations, smaller and more numerous elements are present where there are large variations of stress over small distances (Fig. 4, bottom).

Through optimization, many characteristics of the container can be optimized but only one at a time. This is a problem because we want to minimize many values at the same time, namely the total weight, the maximum stress in each material, the maximum deformation due to a uniform layer of mercury, the maximum deformation due to the displacement of the mercury when the container is tilted, and the maximum deformation when the temperature is reduced from 20°C to -20°C. A quality function has been developed in which all of the above except the weight are variables. This function has been minimized instead of one of the variables. The problem of minimizing the weight is different. For each weight, there is a minimum value of the quality function, but this value will decrease when the weight increases, so there is no clear choice. The weight must be increased until all variables to minimize have a sufficiently small value.

A quality function was first defined for each variable as:

$$Q_x = \frac{1}{\dfrac{L_x}{M_x} - 1} \tag{1}$$

Where $Q_x$ is the quality function of variable x which is a stress or a deformation, $L_x$ is the safe limit for this variable and $M_x$ is the maximum value of this variable on the whole container. The quality function is so that it equal infinity when $M_x = L_x$ and equal 0 when $M_x = 0$. Then the total quality function was defined as the sum of the square of the 6 individual quality functions with their appropriate weighting:

$$Q = Q_F{}^2 + Q_K{}^2 + Q_A{}^2 + 4\,Q_T{}^2 + Q_R{}^2 + Q_M{}^2 \tag{2}$$

Where Q is the total quality function that was minimised in the optimisation process, $Q_F$, $Q_K$ and $Q_A$ are the quality function of the maximum stresses in respectively the foam, the kevlar and the aluminium, $Q_T$ is the quality function of the rigidity in tilt, $Q_R$ is the quality function of the maximum reduction in mercury thickness due to the deformation when the temperature is reduced by 40°C, $Q_M$ is the quality function of the edge deformation when 1 mm of mercury is applied. The maximum stresses were calculated using the maximum thickness of mercury that may be necessary when starting the container rotation (2.5 mm) and a drop of temperature of 40°C. $Q_T{}^2$ was given a weight of 4 because the total rigidity in tilt is not only due to the container rigidity but also to the bearing and base rigidities, so the container must be more rigid than if it was the only contribution. Also, the rigidity is important for other reasons as the resistance to vibrations. $L_R$, the safe limit for the maximum reduction in mercury thickness over the container when the temperature drops 40C, was defined as 0.7 mm. This is for a thickness of 1 mm, which means that 0.3 mm only of mercury would remain over the highest point of the bottom surface. A larger value would easily end in the breaking of the mercury surface.

## 5. Results of the simulations

The main results of the calculations are the dimensions of all the parts of the containers. These are given in appendix 1. The central cylinders used are of 2 diameters, about 300 mm and 500 mm. The former is the largest easily available "on the shelf" while the latter must be custom made as far as we know. All of the values are in reasonnable ranges. For example, the container of 6-m diameter has a thickness of less than 1.2-m. This is reasonnable considering that the foam-kevlar technology is used to build large objects as boats and planes.

Table 1 shows the main characteristics of the resulting containers. The total maximum weight of the 6-m with 2.5 mm of mercury is about 2100 kg, a much smaller weight than a 6-m glass mirror and its metallic sustaining structure. Although the container is expected to operate with a 1-mm mercury thickness, a thickness of the order of 2.5-mm is needed upon startup (Borra, Content & Girard 1993). The aluminum cylinder has been designed with minimum machining in mind. With a diameter of only 50 cm, a length of 110 cm and a total weight of 130 kg, it should be possible to machine it in most workshops. Only both ends need to be machined in order to fit with the aluminum plates, further reducing construction time and cost. Most of the construction time is needed to laminate the surface of Kevlar, a total of 490 m$^2$ of fabrics, but this can be done by unskilled labor.

### 5.1. Constraints

In order to calculate the maximum stresses and deformations, a simulation was performed with the largest loading of mercury (a thickness of 2.5 mm) and the largest temperature variation (from 20°C to -20°C). The results show that all 3 maximum stresses, one per material, are near the edge of one of the 2 aluminum plates (Fig. 5). Along with the stresses and deformations of the container, the program gives an evaluation of the uncertainty on maximum stresses. The maximum stress with uncertainty included can then be calculated for each material and compared with the largest stress that we consider safe. The security factors, defined as stress ratios of the later to the former, are shown in Table 2 for each material. We can see that all values are sufficiently large for the container to be declared safe. The security factors are in fact pessimistic because the main cause of the maximum stresses is the difference between the thermal expansion coefficients, not the weight. The foam sustains a larger contraction than aluminum when there is a reduction in temperature while the Kevlar expands almost only in thickness, with a much smaller contraction in the material plane. Stresses build up from these differences. However, in this case it is the deformation, not the stress, which is the main factor. If the security factors were calculated from the ratios of the largest safe deformation to the maximum deformation, they would be much larger because the materials can become first non-linear then plastic, and therefore expand considerably without breaking. Also, the Kevlar is sustained by the aluminum plate at its point of maximum stress while the largest safe stress has been determined from experiences on Kevlar alone. Finally, the position of maximum stress in aluminium (end of the top plate) is inside the compound kevlar-aluminium encircled by material under much smaller stress. That material would maintain the solidity of the container if the material at this specific end was to fail. This is important because the exact values of the stresses in the glue, kevlar and aluminium at the very ends of the plates is extremely difficult to calculate. An exact description of all 3 materials at a resolution of a few micrometers would be necessary there to get precise values.

## 5.2. Tilt deformations

The security factor for the rigidity in tilt is calculated from the additional tilt generated by the displacement of the mercury when a tilt is applied to the container. Its value is the ratio of the applied tilt to the additional tilt. When a tilt is applied, less mercury will remain at the top of the slope lowering the mercury pressure there while a corresponding increase in weight and pressure will appear at the bottom of the slope. The resulting torque will produce the additionnal tilt. With the previous design, the security ratios were between 1.5 and 3.0 and were limited by the shear deformation of the foam (Fig. 2). In the new design, the rigidity is much larger especially considering that it must grow with the power 4 of the diameter to give the same security factor. The rigidity can also be easily increased if one would accept a higher weight by adding aluminum and Kevlar.

## 5.3. Temperature deformations

Because the mirror is liquid, the container can sustain deformations of many hundreds of wavelengths without significant changes in the optical quality of the mercury surface. The mercury simply moves from the highest point to the lowest, maintaining the shape of the reflecting parabolic surface. However, an upward deformation larger than the mercury thickness would pierce it. This is not as bad a problem as one may think, for the hole will not grow. Figure 6 shows the shape of a 6-m container surface with respect to an ideal parabola for the temperature of construction (+20°C) and the minimum temperature (-20°C). At 20°C, the deformation is due to the weight of the mercury and by the container speed at which the resin was let to solidify. This speed was volontarily chosen slightly slower to equilibrate the height of the highest point at +20°C and –20°C. A hole a few tenths of mm deep near the centre reduces the maximum height at -20°C. At this temperature, the larger contraction of the foam on a 6-m container creates a depression of 0.47 mm at a radius of 0.66 m. The thickness of mercury being larger at that point, the waves on the surface are less damped. However, the damping is larger near the edge of the mirror where the thickness of mercury is smaller. Basically, the total damping is a function of the average thickness, not the lowest point. The situation is however more favorable than suggested by the P-V value, since the large deflection occurs at <1.5 m radius, hence involves <25% of the area of the mirror where the inconvenients of a thicker layer (larger mass, less damping) are not as serious. Note also that, in practice, the mercury layer of a working mirror would be pumped down to 1 mm average so that we would have 1.5-mm thickness at worse, which still give reasonable damping. Also, fortunately, it is far more important to have a substantial damping in the outer regions, which are subject to spiral-shaped wind-induced disturbances (Paper I, Paper II), than in the inner regions which are only affected by vibration-induced ripples which are a minor source of scattered light.

The most important value remains the height of the highest point, which locally reduces the thickness of mercury. We can see that it is never higher than 0.18 mm (4-m with small cylinder), a small value compared to the minimum thickness of 1.00 mm that we are planning to use. On our preliminary design, the height of the highest point at -20°C was at the edge and was much larger because the conical Kevlar layer underneath the container is particularly rigid and was stopping the edge from contracting. As explained in section 3.2, this problem has been resolved by moving some of the Kevlar from the bottom to the top surface of the container. The overall rigidity remains nearly the same but the bottom Kevlar layer sustains a larger contraction than the top when the temperature drops, which allows the edge to contract similarly to the rest of the container.

## 5.3. Measurements in the laboratory

A 3.7-m with 3.6-m clear aperture and a flat 1-m have been built in our laboratory at Laval University and thouroughly tested. The goal was to certify the theoretical calculations by finite elements analysis of the liquid mirror containers in the present paper. The image quality given by the rotating mercury surface of the 3.7-m was also optically tested and is analysed in Tremblay & Borra 2000. We performed a series of measurements on the 3.7-m to determined the deformations when loaded and another series on the 1-m when temperature variations are applied. A companion paper describes in detail these measurements and the comparison with the simulations. The results confirm that the theoretical calculations are in reasonable agreement with the experimental measurements.

The 3.7-m was designed in a similar manner than the containers described in this paper but the kevlar and the cylinder were modeled with "skin" elements which have one less dimension. In these elements, the thickness becomes a parameter. Also, since there was no significant temperature differences in the laboratory, the container was optimized to maximize the rigidity in tilt only (the fourth term at the right of equation 2) and there was no top aluminium plate which role is mostly to reduce the large constaints in the foam at the junction between kevlar and aluminium when the temperature changes. The weight being the only load, the optimisation process gave a larger thickness of kevlar on the bottom of the container, not the top as in the design described in this paper. This is because the conical bottom of the container is better than the parabolic top to sustain the weight and to resist the torque of the mercury caused by a tilt. Figure 7 shows the constraints and (exaggerated) deformations of the foam due to a uniform layer of mercury. The maximum stress is near the cylinder since it supports the weight of the foam and kevlar. We can see how the conic kevlar bottom stops the edge from bending down. Measurements were made after placing bottles of mercury at 1 position for simulation of tilt-like deformations, 2 positions for simulation of saddle shape deformations, and many positions for simulation of a uniform thickness of mercury. The deformations being small and the measurements difficult, I developped a statistical method of analysis that minimizes the statistical uncertainties. The method is described in the companion paper.

The 1-m was specifically designed to measure the vertical deformation of the top surface when the temperature is reduced and compare it with the theory. To make the measurements easier, the top was made flat instead of parabolic and the design was made to maximize the temperature deformations. Since the largest deformation is due to the difference in thermal expansion between the foam and the cylinder, kevlar was used to build the cylinder instead of aluminium because of the larger difference with foam of the kevlar coefficient of thermal expansion. Also, no aluminium plates were put at the top and bottom since their role is to reduce the foam constraints, which would reduce the deformations. The ratio of central thickness to diameter was also increased since deformations are proportionnal to length. This also increases the effect of the conical shape because the angle of the cone is steeper. The angle was chosen to be 45° because a larger angle would make the container too different from the liquid mirror containers it must simulates. Altogether, this made the 1-m P-V (Peak-to-Valley) of the deformation equal to the deformation P-V of a 2.2-m. Skin elements were used for the kevlar in computer simulations except where it was thicker in order to better simulate the real container. Figure 8 shows the constraints and (again exagerated) deformations of the foam with a reduction in temperature. We can see how the lack of a top aluminium plate leaves the foam pinched between the top surface and the cylinder and how the conic bottom stops the edge from contracting. As for the 3.7-m, deformations were small and measurements difficult, especially since they were done by hand down to temperatures of -30°C. Many hundreds of measurements were taken to compensate. For my part, I developed a statistical method based on a model with a large number of calculated coefficients. As for the 3.7-m, this maximizes the statistical precision. The method is described in the companion paper.

## 6. Conclusion

The computer simulations have shown that the new design proposed in this article gives acceptable stresses and deformations of the containers for mirrors up to 6 meters in diameters. Measurements on a 3.7-m and a 1-m flat container (see the companion paper: Borra et al. 2003) show that theory and experiments are in reasonable agreement. The figures and tables contained in this paper can be used as recipes to build containers having diameters between 2 and 6 meters. Table 1 gives the dimensions of the main components. Appendix 1 gives the dimensions and thicknesses of every parts of the containers with figure 3 showing where these parts are positionned. The low cost and simplicity of fabrication of these containers permit the building of an observatory with a large number of large mirrors. The light collecting area of such an observatory would be the same as that of a much larger telescope. For example, an observatory of 40 x 4-m would have the same collecting area than a 25-m but at a cost smaller than one stearable glass 4-m.

## 7. Acknowledgments

This work was supported by Durham University and the Particle Physic & Astronomy Research Council as well as the Natural Sciences and Engeneering research council of Canada. I wish to thank F. Arrien and C. Gosselin for early work on containers. A large number of undergraduate engineering students participated in early simulation work: I thank them all.

## 9. Appendices

### 9.1. Definitions of the containers

Table 3 gives the distances for each values of the containers for each diameter. The names refer to the distances and numbers of kevlar layers in Figure 3.

### 9.2. Rigidity in tilt

When a rotating liquid mirror container is tilted, the mercury will move down the slope if the tilt rotates with the mirror. The bottom of the slope will be filled by mercury coming from the top. The displaced mercury will create a moment on the container that will make it tilt even more. The tilt will also create a moment in the opposite direction due to the angular rigidity of the container. For the container to be angularly stable, the moment of the container must be larger than the moment of the unbalanced mercury. We have:

$$M1 = C\Phi \tag{3}$$

where M1 is the moment created by the system container-bearing-base when tilted, C is the rigidity of the system, and $\Phi$ is the angle of tilt. Note that:

$$1/C = 1/C_1 + 1/C_2 + 1/C_3 \tag{4}$$

where $C_1$ is the container rigidity, $C_2$ is the bearing rigidity and $C_3$ the rigidity of the base. Integrating over the whole surface in polar coordinates (r, θ), the displaced mercury will create a moment of:

$$M_2 = \rho g \int_0^R dr \int_0^{2\pi} (\Phi r^3 \cos^2 \theta) d\theta \tag{5}$$

$$M_2 = \frac{\pi \rho g D^4 \Phi}{64} \tag{6}$$

where ρ is the density of mercury, g the constant of gravity and D the diameter of the mirror. For the container to be stable, we need:

$$M_1 > M_2 \Rightarrow C > \frac{\pi \rho g D^4}{64} \tag{7}$$

the security factor is then 64 C / ($\pi$ $\rho$ g D$^4$). More details can be found in Content 1992.

The preceeding calculations are a first approximation where it is assumed that a displacement of mercury due to a tilt will cause a deformation that is only an additionnal tilt. The real deformation can be any shape of the form, in polar coordinates:

$$d(r,\theta) = \sum_i A_i r^{2i-1} \cos\theta \qquad (8)$$

where i is an index varying from 1 to infinity and A$_i$ a series of coefficients. A complete calculation to any desired precision can be made by assuming that the number of terms in equation 8 is finite and that the reaction pressure of the container caused by the deformation is also of the form of equation 8 with the same finite number of terms "n". We have tilt alone for n = 1 as in the preceding analysis, tilt and coma for n = 2, etc. The system will be unstable if a deformation auto-amplifies because of the additionnal pressure of mercury filling the deformation: the additionnal deformation due to the pressure of mercury filling an original "forced" deformation is larger than that original deformation and has the same shape. This is very similar to a resonnance problem. As for the resonnance, a series of modes will be found. The basic equation is:

$$P_C(d) = KP_{HG}(d) = K\rho g d \qquad \forall (r,\theta) \,;\, K > 0 \qquad (9)$$

where $P_C$ is the reaction pressure of the container for the deformation d, K the constant of proportionnality and $P_{HG}$ is the pressure of the mercury filling that same deformation (or being pushed away leaving a negative pressure). The system is stable if K > 1. As in a resonnance problem, each value of K gives a mode. Here the pressure and deformation are the components along the vertical; we neglect the centrifugal force which only slightly stretches the container. For each elementary deformation of the shape:

$$d_{0i} = r^{2i-1} \cos\theta \tag{10}$$

the pressure function giving this deformation can be calculated using a program of finite element analysis as ANSYS:

$$P_C(d_{0i}) = P_{0i} = \sum_k B_{ik} r^{2k-1} \cos\theta \tag{11}$$

A program as ANSYS will not usually give the $B_{ki}$ directly because it calculates the deformations for a defined pressure while we need the opposite. One can however calculates with the program the deformations for elementary pressure functions $r^{2i-1} \cos\theta$, then makes a best fit using equation 8 as the model for d. The $B_{ik}$ coefficients can then be calculated from the $A_{ik}$ by a matrix inversion. Inserting equations 8 and 11 into 9:

$$P_C(d) = P_C\left(\sum_i A_i r^{2i-1} \cos\theta\right) = \sum_i A_i P_C\left(r^{2i-1} \cos\theta\right) = \sum_i A_i \sum_k B_{ik} r^{2k-1} \cos\theta \tag{12}$$

$$= \sum_k \left( \sum_i A_i B_{ik} \right) r^{2k-1} \cos\theta = \sum_i \left( \sum_k A_k B_{ki} \right) r^{2i-1} \cos\theta = K\rho g \sum_i A_i r^{2i-1} \cos\theta$$

$$\sum_i \left( \left[ \sum_k A_k B_{ki} \right] - K\rho g A_i \right) r^{2i-1} \cos\theta = 0 \qquad\qquad \forall (r,\theta)\,;\, K > 0 \qquad\qquad (13)$$

This can be put in matrix form:

$$\begin{bmatrix} B_{11} - K\rho g & B_{21} & ... & B_{k1} & ... & B_{n1} \\ B_{12} & B_{22} - K\rho g & ... & B_{k2} & ... & B_{n2} \\ ... & ... & ... & ... & ... & ... \\ B_{1i} & B_{2i} & ... & B_{ii} - K\rho g & ... & B_{ni} \\ ... & ... & ... & ... & ... & ... \\ B_{1n} & B_{2n} & ... & B_{kn} & ... & B_{nn} - K\rho g \end{bmatrix} X \begin{bmatrix} A_1 \\ A_2 \\ ... \\ A_i \\ ... \\ A_n \end{bmatrix} = \begin{bmatrix} 0 \\ 0 \\ ... \\ 0 \\ ... \\ 0 \end{bmatrix} \quad (14)$$

which is the same as in a classic resonance problem: the determinant must be equal to zero for a solution to exist. This gives an equation of degree n for K with n solutions although only the real positive values make sense. $K_0$, the lowest value of K, is the fundamental mode. If $K_0 \leq 1$, the system is unstable. Otherwise, $K_0$ is the security factor.

Equation 8 can be generalised as:

$$d(r,\theta) = \sum_i \sum_j A_{ij} r^{2i-1-\text{mod}([j+1]/2)} \cos j\theta \qquad\qquad (15)$$

where mod([j+1]/2) is j+1 modulo 2, which give 0 or 1. However, since a pressure of the form cos jθ will give a deformation of the form cos jθ, the modes will be of the form f(r) cos jθ. Each value of j has its set of modes but we know from experience that the most unstable is by far the tilt-like deformation for which j = 1. It will then give the fundamental mode. For our containers, the deformation due to a tilt mercury pressure is very similar to a tilt. For a 6-m, the residual of a linear fit leaves a peak-to-valley (P-V) that is only 6.5% of the tilt P-V. Inequation 7 is then a very good approximation. To be sure, the security factor of the stability in tilt in table 2 is calculated from a measurement that makes it pessimistic.

**Figure captions**

**Figure 1:** Four different designs of liquid mirror containers. The design of the first large container with its liquid mirror which is a laboratory 1.5-m (top), the simplest possible container design with foam and kevlar (middle top), the container design used in most operational liquid mirrors (middle bottom) and the new design (bottom).

**Figure 2:** Shematic of the 2 types of deformations, bending (or flexure) and shearing. If the ratio of foam to kevlar rigidity is large, bending is dominant and can be reduced by adding kevlar (top). If the ratio is small enough, shearing is dominant and cannot be reduced by adding kevlar (bottom).

**Figure 3:** Shematic view of the new container design. The dimensions for different diameters are given in Appendice 1.

**Figure 4:** Section of the container model used in the ANSYS program (top) and enlargement of the top part near the container centre (bottom). The elements used in the calculations are visible. Their density was increased where the stress is higher.

**Figure 5:** Stresses under maximum load near their respective maximum in Kevlar (top), Aluminum (middle) and foam (bottom).

**Figure 6:** Height of the 6-m top surface with respect to the ideal parabolic shape at 2 temperatures. The highest point is less than 0.15 mm. The average thickness of mercury over the container would then need to be 1.15 mm to maintain a minimum of 1 mm everywhere.

**Figure 7:** Stress (and exagerated deformations) in the foam of the 3.7-m container when loaded with a uniform layer of mercury.

**Figure 8:** Stress (and exagerated deformations) in the foam of the 1-m flat container when temperature is reduced.

Table 1
Container characteristics

| | | | small | large | | | |
|---|---|---|---|---|---|---|---|
| Minimum Temperature (°C) | -20 | -20 | -20 | | -20 | -20 | -10 |
| **Diameter (m)** | | | | | | | |
| Clear aperture | 2.00 | 3.00 | 4.00 | | 5.00 | 6.00 | 6.00 |
| Hg | 2.05 | 3.05 | 4.05 | | 5.05 | 6.05 | 6.05 |
| Container | 2.10 | 3.10 | 4.10 | | 5.10 | 6.10 | 6.10 |
| Central cylinder | | | small | large | | | |
| **Thickness (mm)** | | | | | | | |
| Centre of container | 222 | 448 | 756 | 633 | 872 | 1142 | 1089 |
| Hg (startup) | 2.0 | 2.0 | 2.5 | 2.5 | 2.5 | 2.5 | 2.5 |
| Hg (operation) | 1.0 | 1.0 | 1.0 | 1.0 | 1.0 | 1.0 | 1.0 |
| **Weight** | | | | | | | |
| Foam | 10 | 38 | 111 | 97 | 208 | 392 | 381 |
| Kevlar | 12.5 | 30 | 58 | 59 | 105 | 192 | 204 |
| Al | 10 | 23 | 36 | 70 | 117 | 175 | 175 |
| Resin (10 mm) | 37 | 81 | 141 | 141 | 218 | 313 | 313 |
| Hg (startup+groove) | 96.5 | 208 | 451 | 451 | 696 | 995 | 995 |
| Microsphere filling | 4 | 8 | 14 | 14 | 22 | 31 | 31 |
| Total | 170 | 388 | 811 | 832 | 1366 | 2098 | 2099 |

Table 2
Maximum stresses and deformations

| | | small | large | | | |
|---|---|---|---|---|---|---|
| D clear Aperture (m) | 2.0 | 3.0 | 4.0 | | 5.0 | 6.0 | 6.0 |
| Minimum Temperature (°C) | -20 | -20 | -20 | | -20 | -20 | -10 |
| Cylinder | | | small | large | | | |
| Security coefficient of stress | | | | | | | |
|   Foam | 4.0 | 3.9 | 3.9 | 3.5 | 3.9 | 4.7 | 5.7 |
|   Kevlar | 2.2 | 2.2 | 2.2 | 2.2 | 2.3 | 2.5 | 3.3 |
|   Aluminium | 3.6 | 3.6 | 3.5 | 3.4 | 3.1 | 2.9 | 3.8 |
| Security coefficient of tilt rigidity | 16.9 | 10.0 | 6.0 | 8.6 | 6.3 | 5.2 | 5.5 |
| Maximum reduction of Hg thickness (mm) +20°C-minimum T | 0.14 | 0.14 | 0.18 | 0.16 | 0.16 | 0.15 | 0.13 |
| Edge deformation for 1 mm of Hg | 0.04 | 0.07 | 0.10 | 0.10 | 0.14 | 0.17 | 0.16 |

Table 3
Definitions of the containers

==================================================

| D clear aperture (m) | 2.0 | 3.0 | 4.0 | | 5.0 | 6.0 | 6.0 |
|---|---|---|---|---|---|---|---|
| Focal ratio | 1.5 | 1.5 | 1.5 | | 1.5 | 1.5 | 1.5 |
| Minimum Temperature (°C) | -20 | -20 | -20 | | -20 | -20 | -10 |
| Cylinder | | | small | large | | | |

Distance in mm:

| | | | | | | | |
|---|---|---|---|---|---|---|---|
| H0 | 296.0 | 564.0 | 914.0 | 788.0 | 1066.0 | 1376.0 | 1352.0 |
| HA1 | 8.0 | 11.0 | 14.0 | 13.7 | 11.8 | 10.0 | 10.0 |
| HA2 | 12.0 | 17.0 | 14.0 | 14.3 | 16.2 | 18.0 | 18.0 |
| HA3 | 7.0 | 10.0 | 12.7 | 12.7 | 12.7 | 12.7 | 12.7 |
| HA4 | 33.0 | 0.0 | 0.0 | 0.0 | 0.0 | 0.0 | 0.0 |
| HA5 | 134.3 | 412.6 | 715.1 | 594.2 | 831.2 | 1099.3 | 1069.0 |
| HA6 | 25.0 | 0.0 | 0.0 | 0.0 | 0.0 | 0.0 | 0.0 |
| HA7 | 8.0 | 9.0 | 10.0 | 8.0 | 9.0 | 9.5 | 9.5 |
| HE1 | 50.0 | 50.0 | 50.0 | 50.0 | 50.0 | 50.0 | 50.0 |
| HE2 | 80.0 | 90.0 | 100.0 | 100.0 | 100.0 | 100.0 | 100.0 |
| J1 | 13.0 | 14.0 | 16.0 | 19.0 | 33.0 | 44.0 | 42.0 |
| J2 | 32.0 | 35.0 | 38.0 | 49.0 | 82.0 | 111.0 | 106.0 |
| J3 | 27.0 | 31.0 | 35.0 | 23.0 | 41.0 | 66.0 | 66.0 |
| J4 | 13.0 | 14.0 | 16.0 | 19.0 | 33.0 | 44.0 | 42.0 |
| J5 | 20.0 | 21.0 | 23.0 | 29.0 | 49.0 | 66.0 | 63.0 |
| R0 | 1050.0 | 1550.0 | 2050.0 | 2050.0 | 2550.0 | 3050.0 | 3050.0 |
| RA1 | 239.0 | 254.0 | 265.0 | 370.0 | 454.0 | 537.0 | 527.0 |
| RA2 | 148.0 | 152.4 | 152.4 | 250.0 | 250.0 | 250.0 | 250.0 |
| RA3 | 139.0 | 141.2 | 141.2 | 233.2 | 227.8 | 222.5 | 221.5 |
| RA4 | 141.5 | 152.4 | 152.4 | 250.0 | 250.0 | 250.0 | 250.0 |
| RA5 | 136.5 | 139.7 | 139.7 | 231.7 | 226.3 | 221.0 | 220.0 |
| RA6 | 145.5 | 152.4 | 152.4 | 250.0 | 250.0 | 250.0 | 250.0 |
| RK1 | - | - | - | - | - | 924.0 | 1094.0 |
| RK2 | - | - | - | - | 579.0 | 775.0 | 875.0 |
| RK3 | - | - | 345.0 | 450.0 | 554.0 | 657.0 | 710.0 |
| RK4 | - | 314.0 | 325.0 | 430.0 | 529.0 | 627.0 | 645.0 |
| RK5 | 279.0 | 294.0 | 305.0 | 410.0 | 504.0 | 597.0 | 612.0 |
| RK6 | 259.0 | 274.0 | 285.0 | 390.0 | 479.0 | 567.0 | 557.0 |

Number of layers:

| | | | | | | | |
|---|---|---|---|---|---|---|---|
| HK1 | 6 | 10 | 15 | 15 | 21 | 31 | 31 |

| HK2  | 6  | 10 | 15 | 15 | 21 | 32 | 32 |
|------|----|----|----|----|----|----|----|
| HK3  | 12 | 20 | 30 | 30 | 42 | 63 | 63 |
| HK4  | 8  | 15 | 24 | 24 | 35 | 54 | 54 |
| HK5  | -  | 10 | 18 | 18 | 28 | 45 | 45 |
| HK6  | -  | -  | 12 | 12 | 21 | 36 | 36 |
| HK7  | -  | -  | -  | -  | 14 | 27 | 27 |
| HK8  | -  | -  | -  | -  | -  | 18 | 18 |
| HK9  | 4  | 5  | 6  | 6  | 7  | 9  | 9  |
| HK10 | 6  | 8  | 10 | 10 | 12 | 14 | 14 |
| HK11 | 6  | 8  | 10 | 10 | 12 | 14 | 14 |
| HK12 | 12 | 16 | 20 | 20 | 24 | 28 | 28 |
| HK13 | 8  | 12 | 16 | 16 | 20 | 24 | 24 |
| HK14 | -  | 8  | 12 | 12 | 16 | 20 | 20 |
| HK15 | -  | -  | 8  | 8  | 12 | 16 | 16 |
| HK16 | -  | -  | -  | -  | 8  | 12 | 12 |
| HK17 | -  | -  | -  | -  | -  | 8  | 8  |
| HK18 | 4  | 4  | 4  | 4  | 4  | 4  | 4  |
| HK19 | 2  | 2  | 2  | 2  | 2  | 2  | 2  |

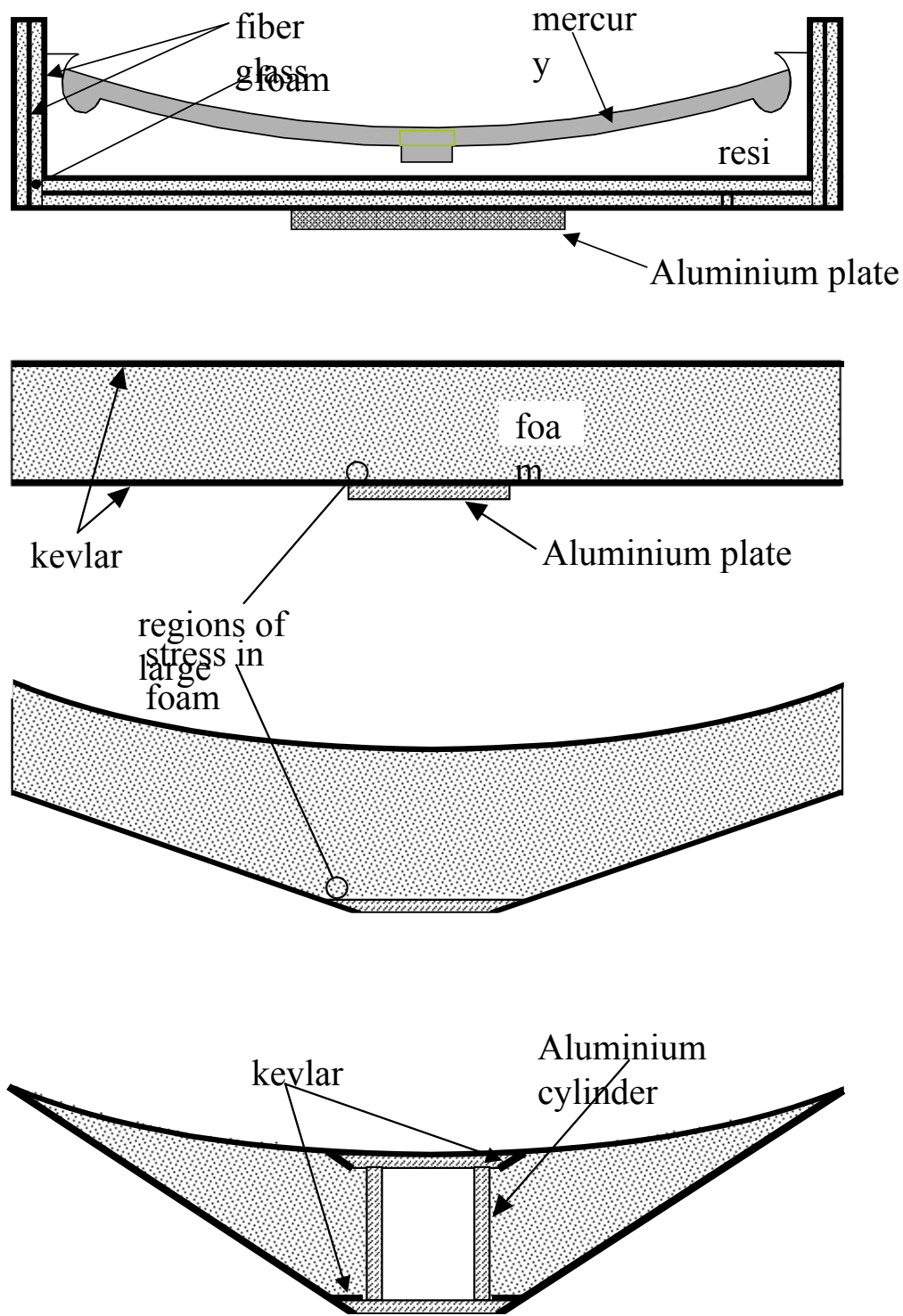

Figure: 1

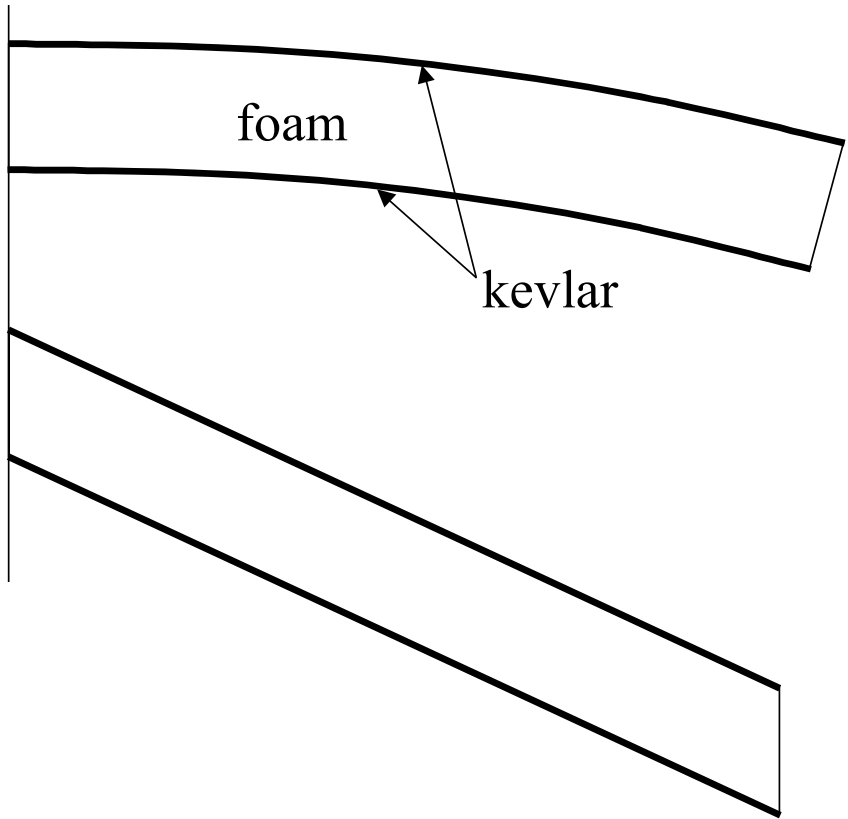

Figure 2

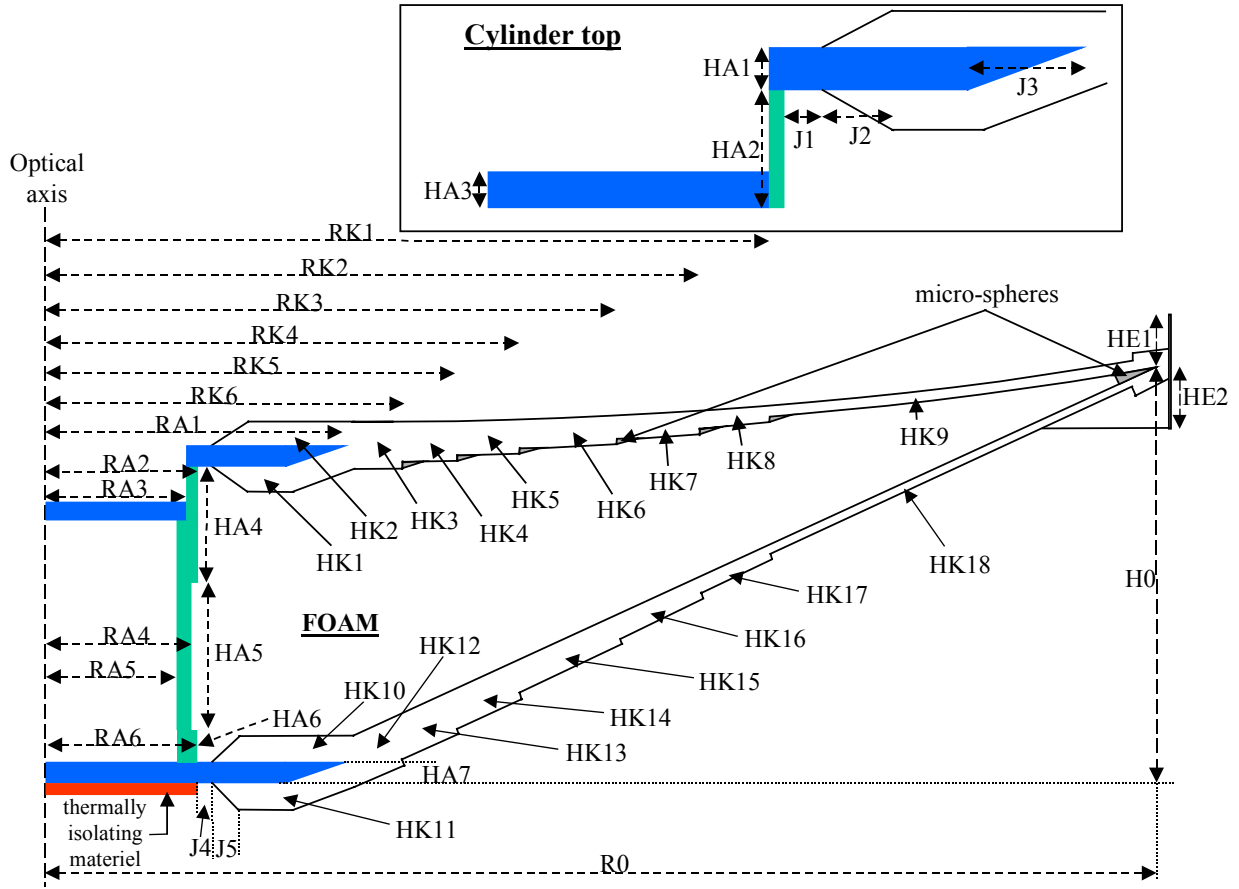

**Figure 3**

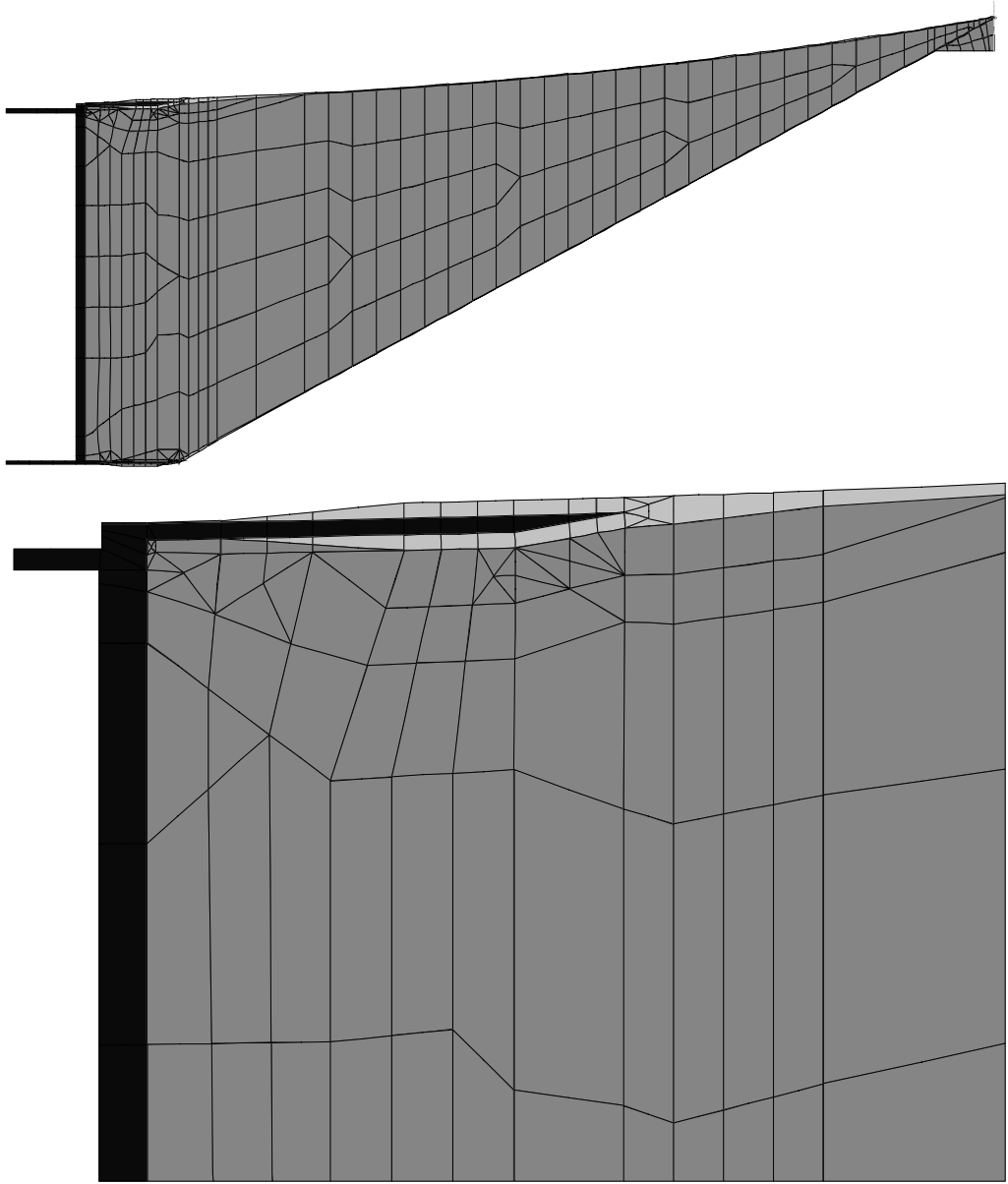

**Figure 4:** Section of the mirror model used in the ANSYS program (top) and enlargement of the top part near the mirror centre (bottom). The elements used in the calculations are visible. Their density was increased where the stress is higher.

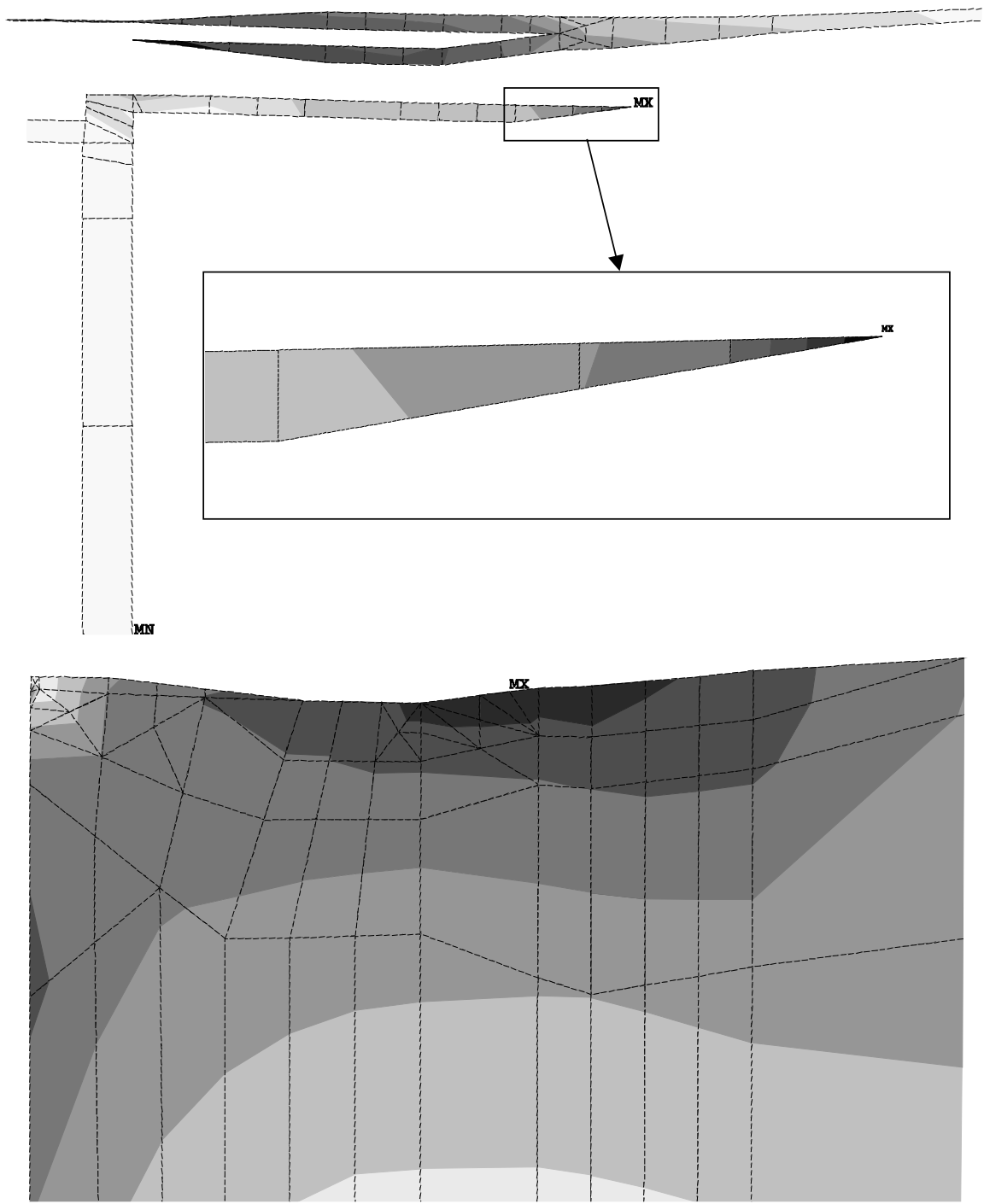

**Figure 5:** Stresses near their respective maximum in Kevlar (top),
Aluminium (middle) and foam (bottom).

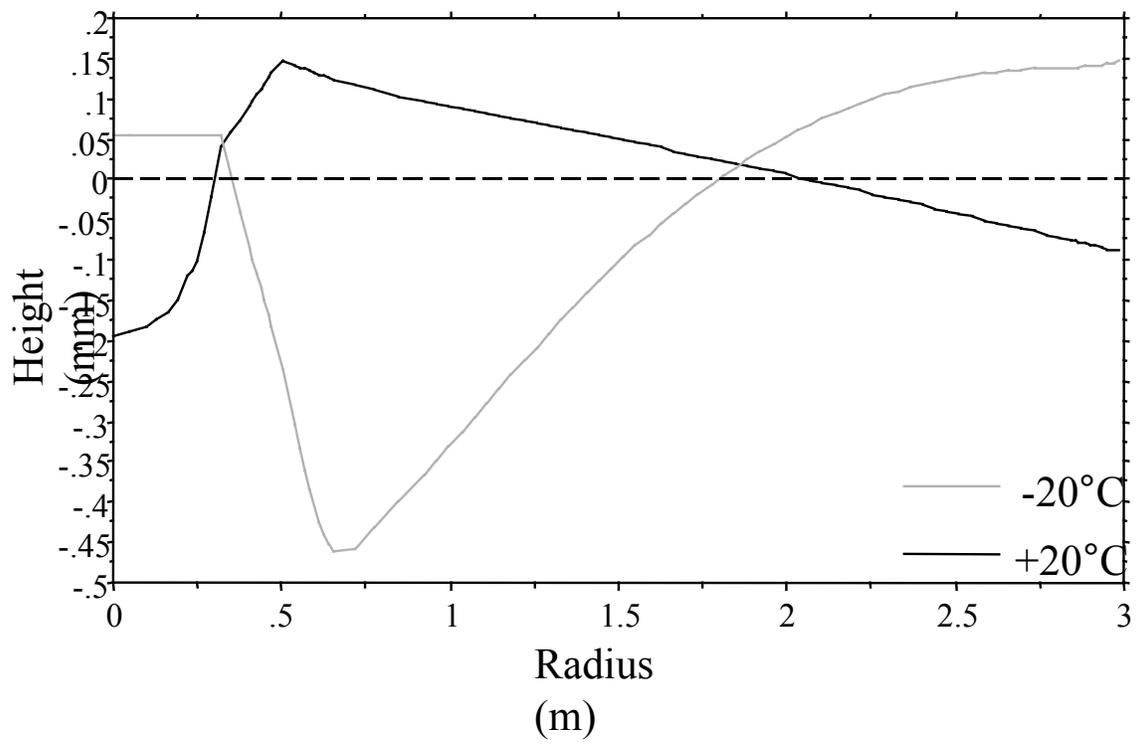

<u>Figure 6</u>

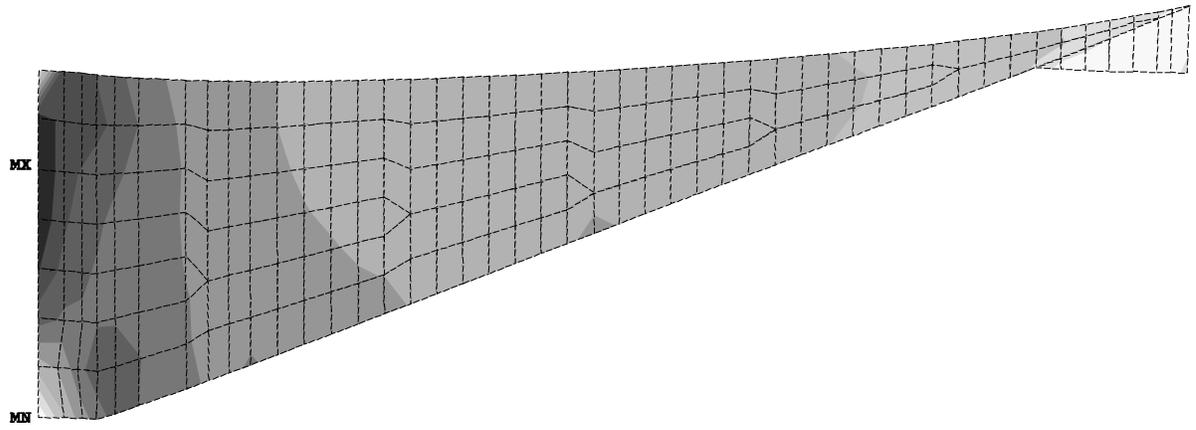

Figure 7